\documentclass{article}
\usepackage{LaThuileFPSpro}
\def\lsim{\lower.7ex\hbox{${\buildrel < \over \sim}$}}
\def\gsim{\lower.7ex\hbox{${\buildrel > \over \sim}$}}
\begin{document}
\title{
  ANISOTROPY OF THE PRIMARY COSMIC-RAY FLUX\\ IN 
  SUPER-KAMIOKANDE\footnote{~~Talk at \char'134 Les Rencontres de Physique de la Vallee
  d'Aosta (La Thuile 2006)'', La Thuile, Aosta Valley, Italy, March 5-11, 2006.}
  \footnote{~~~The talk is based on
  G.Guillian {\it et al.} (Super-Kamiokande collaboration), submitted to Phys.Rev.D, astro-ph/0508468.}
  \footnote{~The PowerPoint file used in the talk can be downloaded from
  {\tt http://www-nu.kek.jp/}\~{\tt oyama/LaThuile.oyama.ppt}}
  }
\author{
  Yuichi Oyama \\
  (for Super-Kamiokande collaboration)\\
  {\em High Energy Accelerator Research Organization (KEK)}\\
  {\em  Oho 1-1 Tsukuba Ibaraki 305-0801, Japan}\\
  E-mail address:~{\tt yuichi.oyama@kek.jp}\\
  URL:~{\tt http://www-nu.kek.jp/}\~{\tt oyama}\\
}
\maketitle

\baselineskip=11.6pt

\begin{abstract}
A first-ever 2-dimensional celestial map of primary cosmic-ray flux was obtained
from $2.10 \times 10^{8}$ cosmic-ray muons accumulated in 1662.0~days of Super-Kamiokande.
The celestial map indicates an $(0.104\pm0.020)$\% excess region in the constellation of Taurus
and a $-(0.094\pm0.014)$\% deficit region toward Virgo.
Interpretations of this anisotropy are discussed.
\end{abstract}
\newpage
\section{Super-Kamiokande detector and cosmic-ray muon data}
Super-Kamiokande (SK) is a large imaging water Cherenkov detector
located at $\sim$2400~m.w.e. underground in the Kamioka mine, Japan.
The geographical coordinates are 36.43$^{\circ}$N latitude
and 137.31$^{\circ}$E longitude.
Fifty~ktons of water in a cylindrical tank is viewed by 11146
20-inch$\phi$ photomultipliers.

The main purpose of the SK experiment is neutrino physics.
In fact, SK has reported many successful results on
atmospheric neutrinos and on solar neutrinos. For recent
results on neutrino physics as well as the present status of
the SK detector, see Koshio.\cite{Koshio}

The SK detector records cosmic-ray muons with an average
rate of $\sim$1.77~Hz.
Because of more than a 2400~m.w.e. rock overburden,
muons with energy larger than $\sim$1~TeV at the ground level can reach
the SK detector. The median energy of parent cosmic-ray primary protons
(and heavier nuclei) for 1~TeV muon is $\sim$10~TeV.

Cosmic-ray muons between June 1, 1996 and May 31, 2001 were used in the
following reported analysis.
The detector live time was 1662.0~days, which corresponds
to a 91.0\% live time fraction.
The number of cosmic-ray muons during this period was $2.54\times 10^{8}$
from 1000 m$^{2}\sim$1200 m$^{2}$ of detection area.

Muon track reconstructions were performed with the standard muon fit algorithm,
which was developed to examine the spatial correlation with spallation
products in solar neutrino analysis.\cite{Ishino}
In order to maintain an angular resolution within 2$^{\circ}$,
muons were required to have track length in the detector greater than 10~m
and be downward-going.
The total number of muon events after these cuts was $2.10\times 10^{8}$,
corresponding to an efficiency of 82.6\%.   

\section{Data analysis and results}
\begin{figure}[b!]
\vspace{8.0cm}
\includegraphics{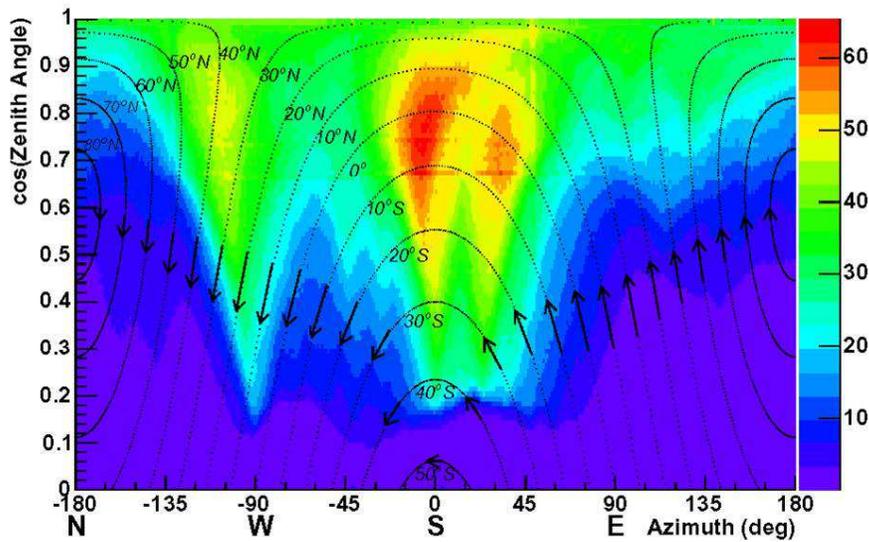}
\caption{
Cosmic-ray muon rate in the horizontal coordinate. The units are ${\rm day^{-1}m^{-2}sr^{-1}}$.
The dot curves indicate contours of constant declination, while the arrows
indicate the apparent motion of stars with the rotation of the Earth.
\label{fig:horizontal}}
\end{figure}

The muon event rate in the horizontal coordinate is shown in Fig.\ref{fig:horizontal}.
The rate is almost constant and the time variation
is less than 1\%.
This distribution merely reflects the shape of the mountain above
the SK detector. For example, the muon flux from the south is larger because
the rock overburden is small in the south direction.

With the rotation of the Earth, a fixed direction in the horizontal
coordinate moves on the celestial sphere.
Therefore, the time variation of muon flux
can be interpreted as the anisotropy of
primary cosmic-ray flux in the celestial coordinate.\cite{celest} 
A fixed direction in the horizontal coordinate travels on a constant declination,
and returns to the same right ascension after one sidereal day.
The muon flux from a given celestial position can be
directly compared with the average flux for the same declination.

\begin{figure}[p!]
\vspace{8.0cm}
\includegraphics{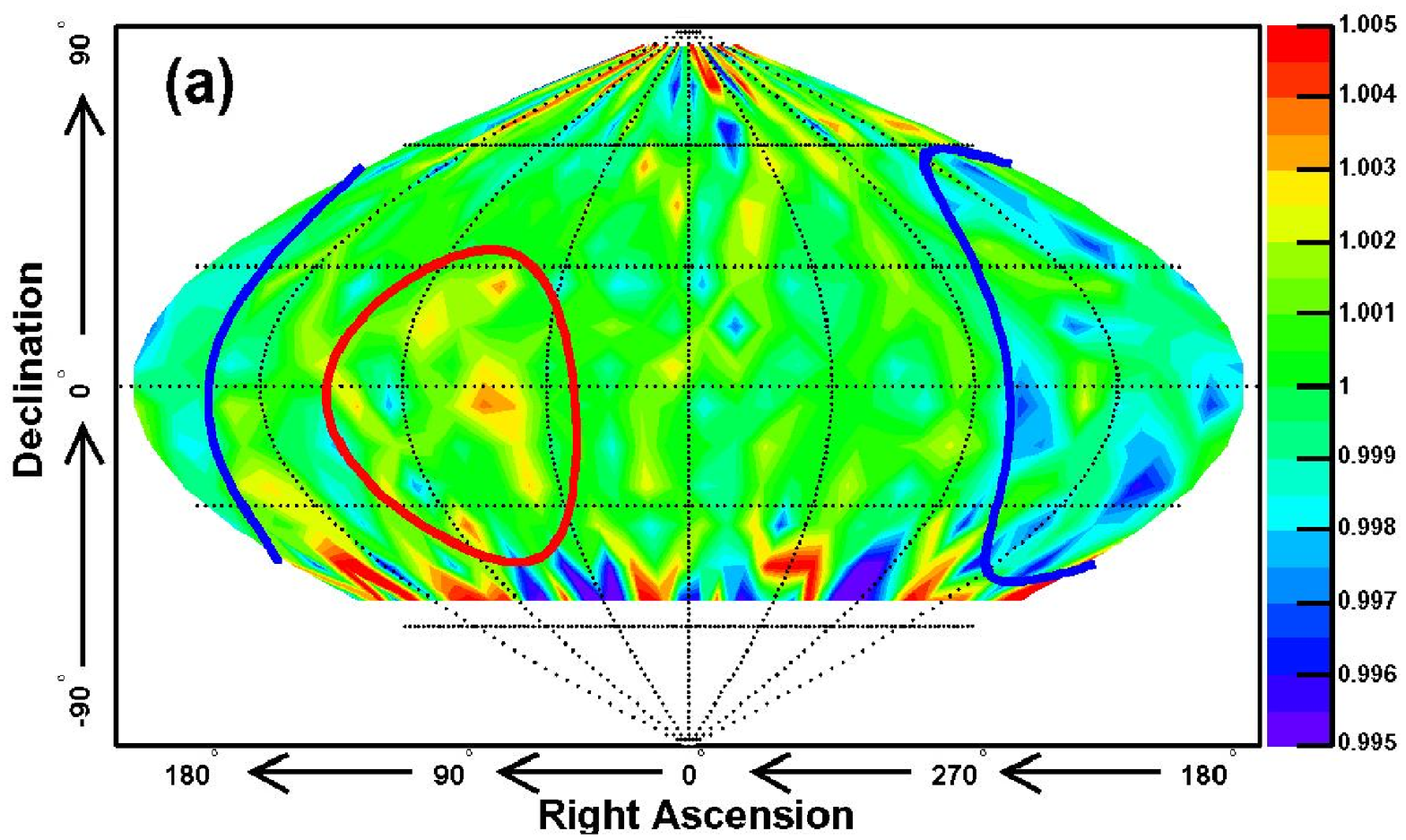}
\vspace{8.0cm}
\includegraphics{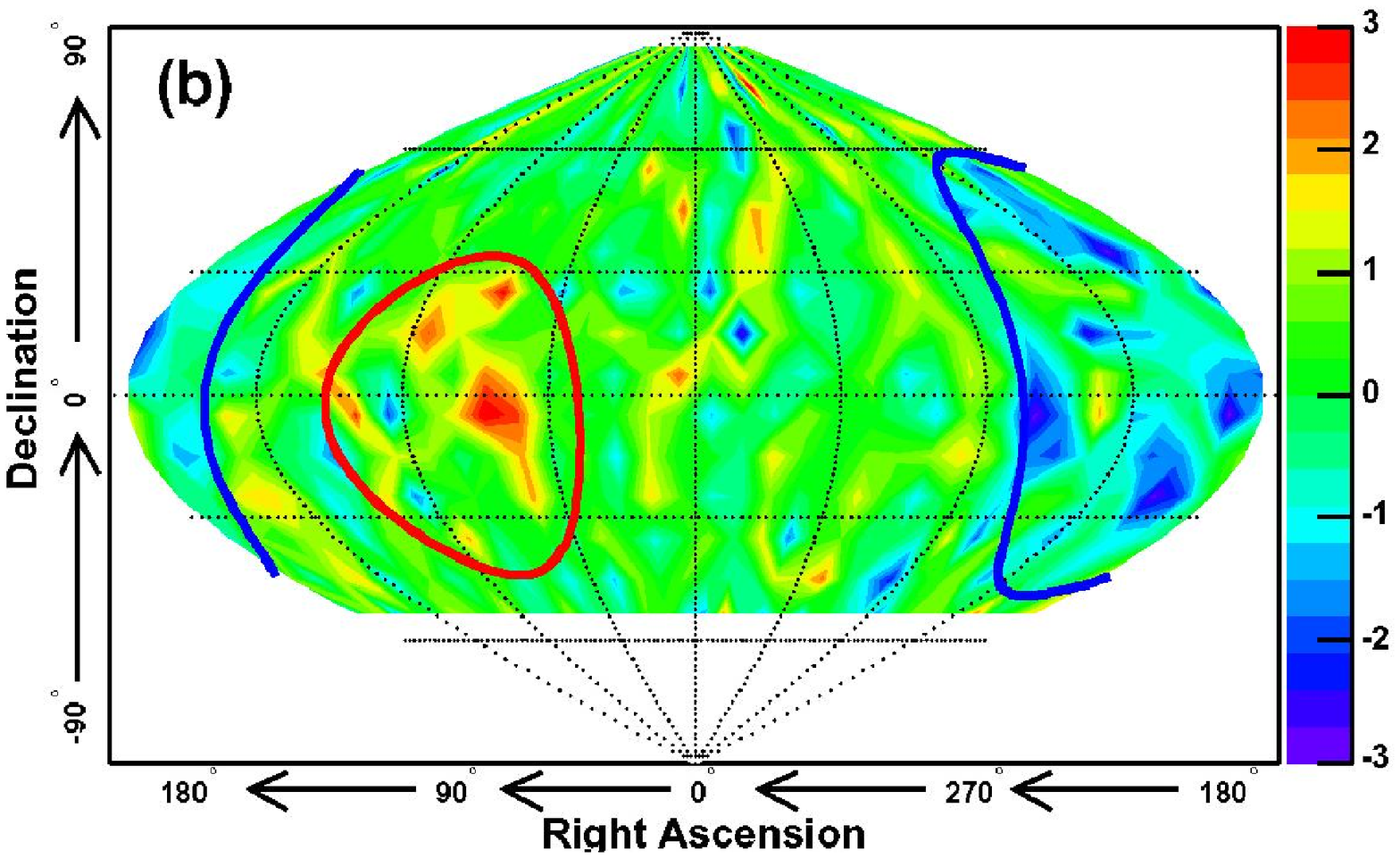}
\caption{
Primary cosmic-ray flux in the celestial coordinate.
Deviations from the average value for the same declinations are shown.
The units are (a)amplitude (from $-$0.5\% to 0.5\%)
and (b)significance (from $-3\sigma$ to $3\sigma$).
The Taurus excess is shown by the  red solid line and the Virgo deficit is
shown by the blue solid line.  
\label{fig:sk2d}}
\end{figure}

Since $360^{\circ}$ of right ascension is viewed in one sidereal day,
the right-ascension distribution is equivalent to the time variation of one sidereal day period.
The cosmic-ray muon flux may have other time variations irrelevant of the celestial anisotropy,
for example,
a change of the upper atmospheric temperature,\cite{Kam2temp} or the orbital motion of the Earth
around the Sun. An interference of one day variation and one-year variation may
produce a fake one sidereal day variation. Those background time variations
are carefully examined and removed to extract $\sim 0.1$\% of the real primary
cosmic-ray anisotropy.
For more details, see G.Guillian {\it et al}.\cite{Guillian}

The deviations of the muon flux from the average for the same declination are shown in
Fig.~\ref{fig:sk2d}.
The units are amplitude in Fig.~\ref{fig:sk2d}(a) and significance in Fig.~\ref{fig:sk2d}(b).
Obviously, an excess is found around
$\alpha \approx 90^{\circ}$ and an deficit around $\alpha \approx 200^{\circ}$.
(The excess and deficit around $\delta \gsim 70^{\circ}$ and
$\delta \lsim -40^{\circ}$ in Fig.\ref{fig:sk2d}(a) are due to poor statistics,
as can be recognized from Fig.~\ref{fig:sk2d}(b).) 

To evaluate the excess and deficit more quantitatively, conical angular windows are defined
with the central position in the celestial coordinates
$(\alpha,\delta)$ and the angular radius, $\Delta\theta$.
If the number of muon events in the angular window is larger or smaller than the average
by 4 standard deviations (which corresponds to chance probability of $6.3\times 10^{-5}$),
the angular window is defined as the excess window or the deficit window.
The celestial position $(\alpha,\delta)$ and
the angular radius ($\Delta\theta$) are adjusted to maximize the statistical significance.

By this method, one significant excess and one significant deficit are found.
From the constellation of their directions, they are named
the Taurus excess and the Virgo deficit. Summary of the Taurus excess and the Virgo deficit
are listed in Table~1. The positions of the Taurus excess and the Virgo deficit are
also shown in Fig.~\ref{fig:sk2d}. 

\begin{table}[t]
\centering
\caption{Amplitude, center of the conical angular windows in the celestial coordinate,
angular radius of the window, the chance probability of finding
the excess or deficit are listed. 
Small chance probabilities might occur somewhere on the map because all positions on
the celestial sphere are surveyed.
Chance probabilities considering such a \char'134 trial factor'' are
also listed in the last row.
}
\vskip 0.1 in
\begin{tabular}{l|c|c}
\hline
\hline
Name   &  Taurus Excess & Virgo deficit \\
\hline
Amplitude   &   $(1.04\pm0.20)\times 10^{-3}$  &  $-(0.94\pm0.14)\times 10^{-3}$ \\  
Center $(\alpha,\delta)$ & $( 75^{\circ}\pm 7^{\circ},-5^{\circ}\pm 9^{\circ})$
                        & $(205^{\circ}\pm 7^{\circ},5^{\circ}\pm 10^{\circ})$ \\
Angular radius $(\Delta\theta)$   & $39^{\circ}\pm 7^{\circ}$  &  $54^{\circ}\pm 7^{\circ}$ \\  
Chance probability      & $2.0\times 10^{-7}$ & $2.1\times 10^{-11}$ \\
~~(trial factor is considered)  & $5.1\times 10^{-6}$ & $7.0\times 10^{-11}$ \\
\hline
\hline
\end{tabular}
\label{tab:excess_deficit}
\end{table}

\section{Comparison with other experiments}

\begin{figure}[p!]
\vspace{12.0cm}
\includegraphics{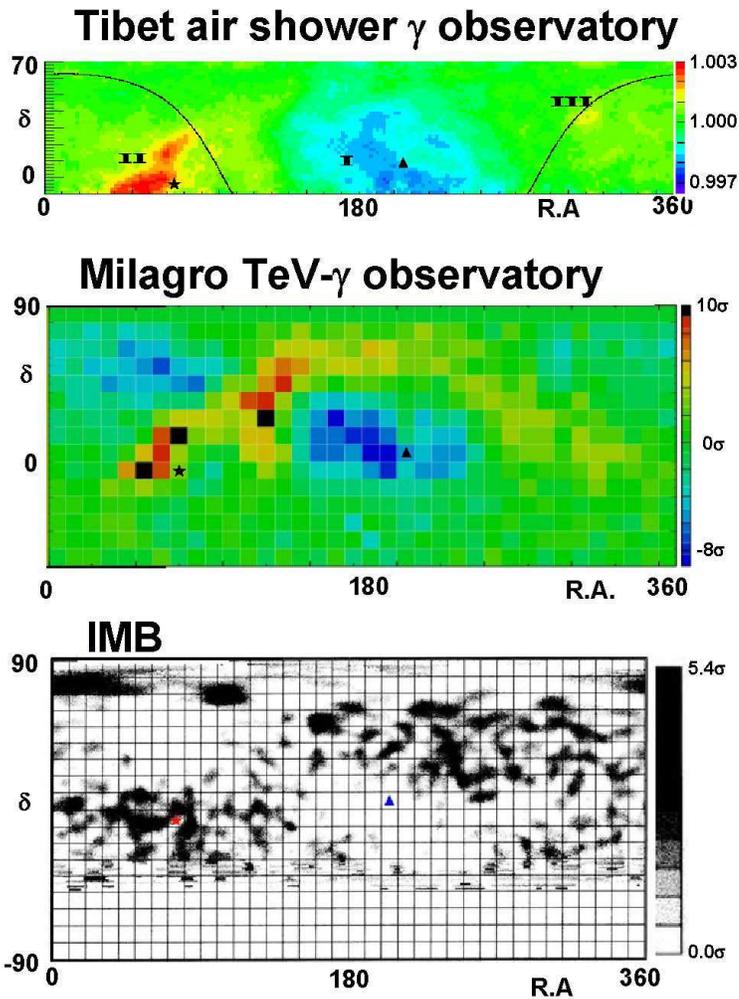}
\caption{
Primary cosmic-ray flux distribution from 3 experiments.
They are Tibet air-shower $\gamma$ observatory~(top),\cite{TibetAS}
Milagro TeV-$\gamma$ observatory~(middle),\cite{Milagro} and
IMB proton decay experiment~(bottom).\cite{IMB}
The center of the Taurus excess by SK is indicated by a star,
and the center of Virgo deficit is indicated by a triangle.
\label{fig:otherexp}}
\end{figure}

Fig.~\ref{fig:sk2d} is the first celestial map of cosmic-ray primaries obtained
from underground muon data. However, there are three similar celestial maps
from other experiments, even though they are not published in any refereed
papers.
Two of them are from $\gamma$-ray observatories: Tibet air shower
$\gamma$ observatory\cite{TibetAS} and Milagro TeV-$\gamma$ observatory.\cite{Milagro}
Note that their primary particles include not only protons, but also $\gamma$-rays, because
of their poor proton/$\gamma$ separation capability.
The other is a celestial map from the IMB proton decay experiment.\cite{IMB}
In the IMB map, only excess regions are plotted.
Results from the 3 experiments are shown in Fig.~\ref{fig:otherexp}.
The trends of 3 celestial maps well agree with the results from SK.

\medskip

In addition to three celestial maps, there were many one-dimensional results from
underground cosmic-ray muon observatories.
Most of the experiments use very simple detectors,
such as 2 or 3 layers of plastic scintillators.
They count cosmic-ray muon rate with coincidence of the plastic scintillator layers. 
All cosmic-ray muons are assumed to arrive from the zenith,
and right-ascension distributions are fitted with first harmonics.
The declination distribution cannot be analyzed.

\begin{figure}[b!]
\vspace{8.0cm}
\includegraphics{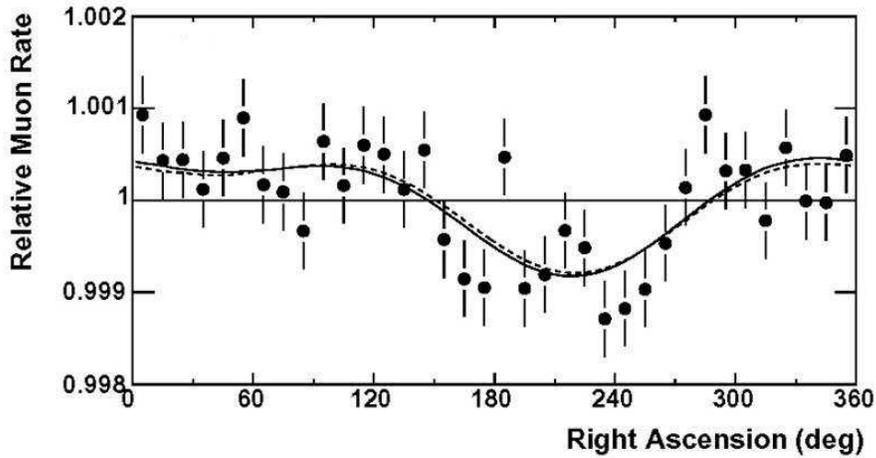}
\caption{
Cosmic-ray muon rate as a function of the right ascension in Super-Kamiokande.
The average muon rate is normalized to be 1.
It is assumed that all muons come from the zenith.
The solid curve is the best fit of the first two harmonic functions.
The dashed curve is the first two harmonics after subtracting the atmospheric
contribution (See Guillian {\it et al.}\cite{Guillian}).
The amplitude and phase of the first harmonics are $(5.3\pm 1.2)\times 10^{-4}$ and
$40^{\circ}\pm 14^{\circ}$. 
\label{fig:sk1dm}}
\end{figure}

\begin{figure}[p!]
\vspace{8.0cm}
\includegraphics{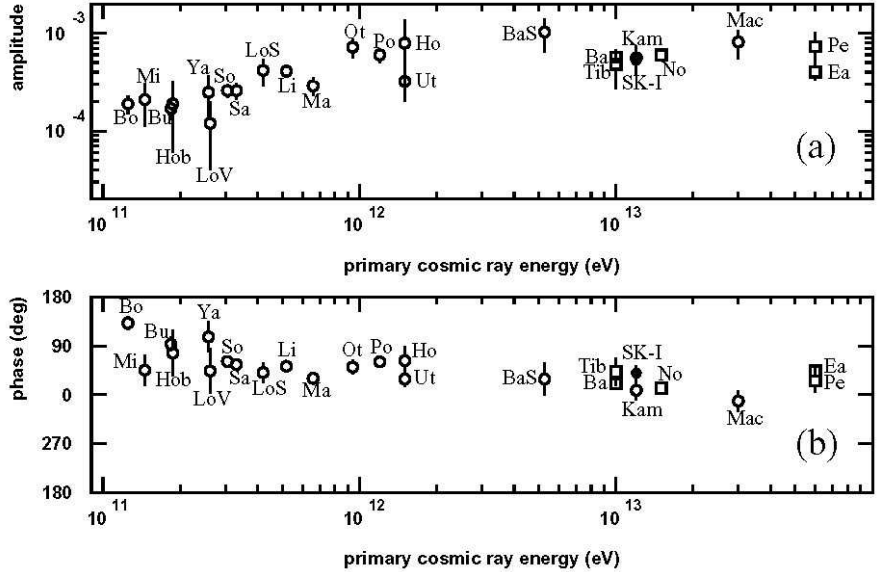}
\caption{
First-harmonic fit of right ascension distributions by various cosmic ray experiments.
The amplitude (top) and phase (bottom) are plotted as a function of the primary energies.
The circles are for underground muon experiments and squares are for extensive air shower arrays.
The filled circle is for Super-Kamiokande. Data references are as follows:
Bo:Bolivia(vertical),\cite{Bolivia} Mi:Misato(vertical),\cite{Misato} Bu:Budapest,\cite{Misato}
Hob:Hobart(vertical),\cite{Misato} Ya:Yakutsk,\cite{Misato} LoV:London(vertical),\cite{Misato}
So:Socomo(vertical),\cite{Bolivia} Sa:Sakashita(vertical),\cite{Sakashita}
LoS:London(south),\cite{London} Li:Liapootah(vertical),\cite{Liapoo}
Ma:Matsushiro(vertical),\cite{Matsushiro} Ot:Ottawa(south),\cite{Ottawa}
Po:Poatina(vertical),\cite{Poatina} Ho:Hong~Kong,\cite{HongKong} Ut:Utah,\cite{Utah}
BaS:Baksan(south),\cite{Baksan} Kam:Kamiokande,\cite{Kam2} Mac:MACRO,\cite{Macro}
Tib:Tibet(vertical),\cite{Tibet} Ba:Baksan~air~shower,\cite{Bakas} No:Mt.~Norikura,\cite{Norikura}
Ea:EAS-TOP,\cite{Eastop} Pe:Peak~Musala.\cite{Musala}
\char'134 (vertical)'' means that the upper plastic scintillator layers are placed exactly
above the bottom layers and the coincidence is sensitive to muons from the zenith.
\char'134 (south)'' means that the upper layers are placed rather
south of the bottom layers and muons from south direction are selectively counted. 
\label{fig:all1dm}}
\end{figure}

Exactly the same analysis method was applied to the SK data to examine
the consistency. The right-ascension distribution is shown in Fig.~\ref{fig:sk1dm}.
The amplitude and the phase of the first harmonics were obtained to be $(5.3\pm 1.2)\times 10^{-4}$ and
$40^{\circ}\pm 14^{\circ}$. 
Results of the analysis are plotted together with other underground muon
experiments and some air shower array experiments in Fig.~\ref{fig:all1dm}.
The agreement with other experiments is excellent.
Especially, the phases of most experiments range between 0$^{\circ}$ and 90$^{\circ}$.

\section{Can protons be used in astronomy?}
Before interpretations of the SK cosmic-ray anisotropy,
trajectories of protons in the galactic magnetic field must be addressed.
The travel directions of protons are bent by the galactic magnetic field in
the Milky Way, which is known to be $\sim 3\times 10^{-10}$~Tesla.
If the direction of the magnetic field is vertical to the proton
direction, the radius of curvature for 10~TeV protons is $\sim 3\times 10^{-3}$~pc.

Since the radius of the solar system is $\sim 2\times 10^{-4}$~pc, 10~TeV protons
keep their directions from outside of the solar system. On the other hand,
since the radius of the Milky Way galaxy is $\sim 20000$~pc, protons may loose their
directions on the scale of the galaxy.

However, if the magnetic field is not vertical to the proton direction,
the trajectories of protons in a uniform magnetic field become spiral. 
The momentum component parallel to the magnetic field remains
after a long travel distance.
Since the galactic magnetic field is thought to be uniform on the order of
$\gsim$300~pc, protons may keep their directions within this scale.
The actual reach of the directional astronomy by protons is unknown.

\section{Excess/deficit and Milky Way galaxy}
\begin{figure}[p!]
\vspace{7.0cm}
\includegraphics{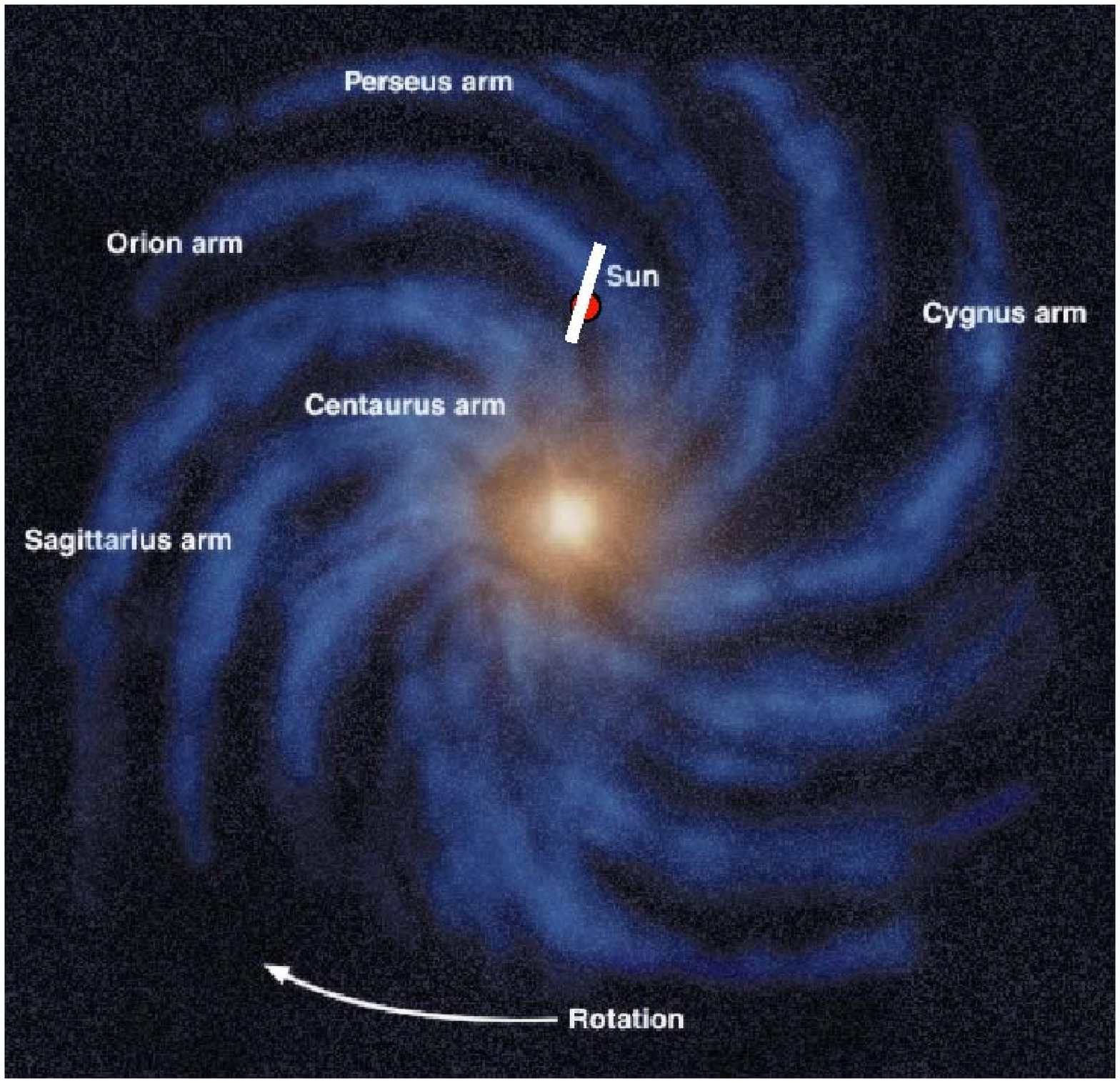}
\vspace{4.0cm}
\includegraphics{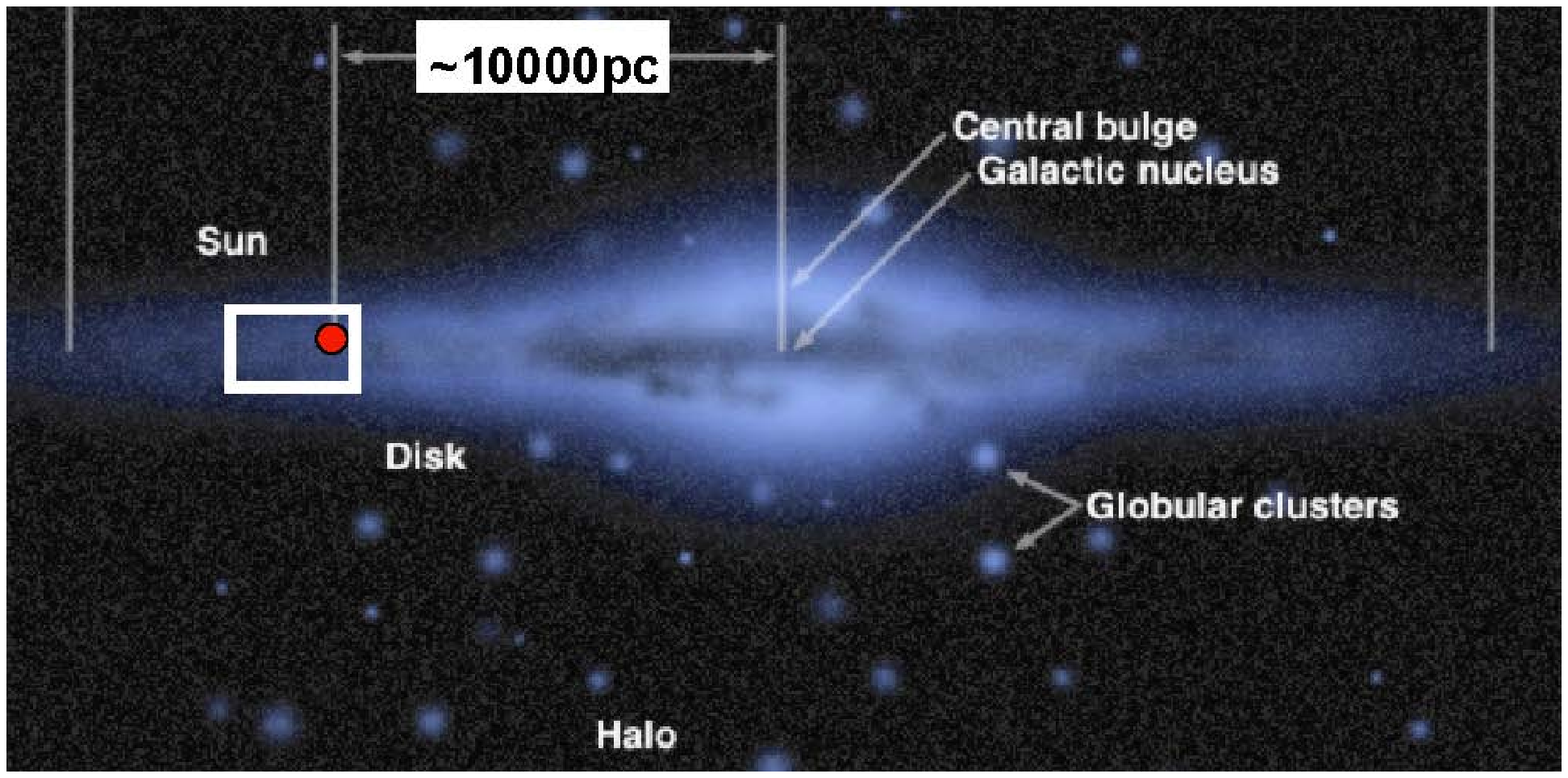}
\vspace{4.0cm}
\includegraphics{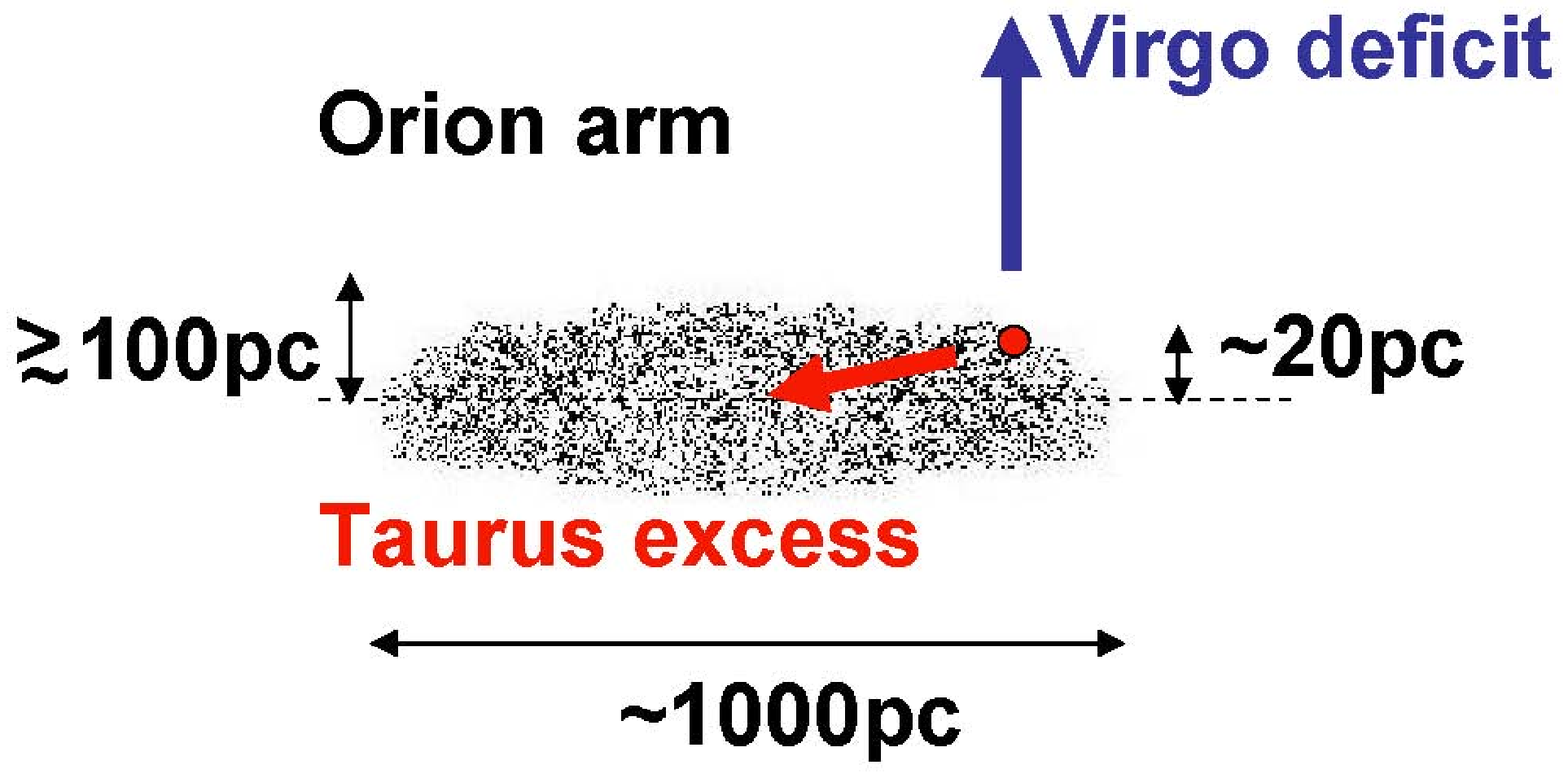}
\caption{
Top view (top) and side view (middle) of the Milky Way galaxy.
The position of the solar system is shown by red circles. A cross-sectional
view of the Orion arm around the Earth is also shown~(bottom).
The Orion arm is $\sim$1000~pc in width and $\gsim$200~pc in thickness.
The solar system is inside of the Orion arm and
$\sim$20~pc away from the center of the Galactic plane. 
The direction of the Taurus excess and the Virgo deficit are also shown.
\label{fig:milkyway}}
\end{figure}

Directional correlations of Taurus excess and Virgo deficit with the Milky Way
galaxy is of great interest.
Schematic illustrations of Milky Way galaxy are shown in Fig.\ref{fig:milkyway}.
Milky Way is a spiral galaxy with 20000~pc radius and $\gsim$200~pc thickness.
The solar system is located about 10000~pc away from the center of the galaxy.
It is in the inside of the Orion arm and about 20~pc away from the galactic
plane, as shown in Fig.\ref{fig:milkyway}(bottom).

The Taurus excess is toward the center of the Orion arm, and the Virgo deficit
is toward the opposite to the galactic plane.
Accordingly, primary cosmic ray flux have a positive correlation with density
of nearby stars around the Orion arm.

\section{Compton-Getting effect}

Assume that  \char'134 cosmic-ray rest system'' exists, in which the cosmic-ray
flux is isotropic. If an observer is moving in this rest system, the cosmic-ray flux
from the forward direction becomes larger. The flux distribution ($\Phi(\theta)$)
shows a dipole structure, which is written as $\Phi(\theta)\propto 1+A\cos\theta$,
where $\theta$ is the angle between the direction of the observer's motion and
the direction of the cosmic-ray flux. 
Such an anisotropy is called the Compton-Getting effect.\cite{CGeffect}
The velocity of the observer ($v$) is proportional
to $A$. If $v$ is 100~km/s, $A$ is $1.6\times 10^{-3}$.

If the Taurus excess ($1.04\times 10^{-3}$) and the Virgo deficit
($-0.94\times 10^{-3}$) were in opposite directions, it might be explained by
the Compton-Getting effect of $v=50\sim 100$km/s.
However, the angular difference between the Taurus excess and the Virgo deficit is about
130$^{\circ}$. The Taurus-Virgo pair is difficult to be explained
by the Compton-Getting effect.
Accordingly, a clear Compton-Getting effect is absent in the SK celestial map.
Although it is difficult to set an upper limit on the relative velocity
because there exist excess and deficit irrelevant to Compton-Getting effect,
it would be safe enough to conclude that the relative velocity
is less than several ten km/s.

The relative velocity between the solar system and the Galactic center is about
200~km/s. The velocity between the solar system and the microwave background is
about 400~km/s.\cite{microwave} The velocity between the Milky Way and the Great Attractor is about
600~km/s.\cite{Greatatt} The upper limit, several ten km/s, is much smaller than those numbers.
The cosmic-ray rest system is not together with the Galactic Center nor
the microwave background nor the Great Attractor, but together with our motion.

Because of the principal of the SK data analysis, two possibilities
cannot be excluded: the Compton-Getting effect is canceled with some
other excess or deficit, and the direction of the observer's motion
is toward $\delta\sim 90^{\circ}$ or $\delta\sim -90^{\circ}$.
\section{Crab pulsar}
One strong interest concerns the correlation of the Taurus excess with
the Crab pulsar. This provides a clue to examine whether high-energy
cosmic rays are accelerated by supernovae or not, which is the most
fundamental problem in cosmic-ray physics.

The Crab pulser\cite{Crab} is a neutron star in the Crab Nebula,
which is one of the closest and newly exploded supernova remnants.
The distance from the solar system is about 2000~pc, and the supernova explosion was in 1054.
The celestial position is $(\alpha,\delta)=(83.63^{\circ},22.02^{\circ})$.
It is in the constellation Taurus and also within the Taurus excess, but is deviated
from the center of the Taurus excess by $28^{\circ}$.

The total energy release from the Crab pulsar is calculated from the
spin-down of the pulsar, and is $4.5\times 10^{38}$erg$\cdot$s$^{-1}$. 
If it is assumed that all energy release goes to the acceleration
of protons up to 10~TeV and the emission of protons is isotropic,
the proton flux at the Earth is calculated to be
${\rm \sim 0.6\times 10^{-7}cm^{-2}s^{-1}}$ 

On the other hand, the Taurus excess observed in SK is converted to
the primary proton flux at the surface of the Earth using the observation
period and the detection area.
The flux is obtained to be ${\rm \sim 1.8\times 10^{-7}cm^{-2}s^{-1}}$

From a comparison of these two numbers,
the Taurus excess cannot be explained by the proton flux accelerated by Crab pulsar
by a factor of $\sim 3$.
Note that two extremely optimistic assumption were implicitly made in calculating
the expected flux from Crab pulsar;
all energy releases are provided to the acceleration of protons
up to 10~TeV, and protons travel straight to the Earth.  
\section{Summary}
The first-ever celestial map of primary cosmic rays ($>$ 10~TeV) was obtained from
$2.10\times 10^{8}$ cosmic-ray muons accumulated in 1662.0~days
of Super-Kamiokande between June 1, 1996 and May 31, 2001.
In the celestial map, one excess and
one deficit are found. 
They are $(1.04\pm 0.20)\times 10^{-3}$ excess from Taurus (Taurus excess) and  
$-(0.94\pm 0.14)\times 10^{-3}$ deficit from Virgo (Virgo deficit).
Both of them are statistically significant.  
Their directions agree well with the density of nearby stars
around the Orion arm. A clear Compton-Getting effect is not found, and the cosmic-ray
rest system is together with our motion. The Taurus excess is difficult to be
explained by the Crab pulsar.

\bigskip

In 1987, Kamiokande started new astronomy beyond \char'134 lights''.
In 2005, Super-Kamiokande started new astronomy
beyond \char'134 neutral particles''.

\bigskip\noindent
{\it -Note Added-}

After the submission of the first draft,\cite{Guillian} it was pointed out that an interpretation
as Compton-Getting effect is not impossible.
Although the angle between the centers of the Taurus excess
and the Virgo deficit is 130$^{\circ}$,
agreement with the dipole structure is fair (not good) because of the quite
large angular radius of the window ($39^{\circ}$ for the Taurus excess and $54^{\circ}$
for the Virgo deficit).
Even if the Taurus-Virgo pair is due to the Compton-Getting effect, the relative velocity
is about 50~km/s. This does not change the discussion about the comparison with
other relative velocities. 

\begin{thebibliography}{99}
\bibitem{celest}As geographical position on the Earth is expressed by longitude and
                latitude, position on the celestial sphere is expressed
                by right ascension $(\alpha)$ and declination $(\delta)$.
                The definition of these parameters can be
                found in any elementary textbook of astronomy.
\end{thebibliography}
\end{document}